# DeepBessel: deep learning-based full-field vibration profilometry using single-shot time-averaged interference microscopy

Maria Cywińska,* Wiktor Forjasz, Emilia Wdowiak, Michał Józwik, Adam Styk, Krzysztof Patorski, and Maciej Trusiak

*Warsaw University of Technology, Institute of Micromechanics and Photonics, 8 Sw. A. Boboli St., 02-525 Warsaw, Poland*
*\*maria.cywinska@pw.edu.pl*

**Abstract:** Full-field vibration profilometry is essential for dynamic characterizing microelectromechanical systems (MEMS/MOEMS). Time-averaged interferometry (TAI) encodes spatial information about MEMS/MOEMS vibration amplitude in the interferogram's amplitude modulation using Bessel function (besselogram). Classical approaches for interferogram analysis are specialized for cosine function fringe patterns and therefore introduce reconstruction errors for besselogram decoding. This paper presents the DeepBessel: a deep learning-based approach for single-shot TAI interferogram analysis. A convolutional neural network (CNN) was trained using synthetic data, where the input consisted of besselograms, and the output corresponded to the underlying vibration amplitude distribution. Numerical validation and experimental testing demonstrated that DeepBessel significantly reduces reconstruction errors compared to the state-of-the-art approaches, e.g., Hilbert Spiral Transform (HST) method. The proposed network effectively mitigates errors caused by the mismatch between the Bessel and cosine functions. The results indicate that deep learning can improve the accuracy of full-field vibration measurements, offering new possibilities for optical metrology in MEMS/MOEMS applications.

## 1. Introduction

Full-field vibration profilometry is a relevant issue in the case of microelectromechanical system (MEMS or MOEMS) manufacturing. When micromechanical component (i.e., micromembrane or microcantilever) is excited to harmonic oscillatory motion, its vibration amplitude distribution can provide the information about material properties in microscale. Vibration analysis can also help with validation of the design and simulation or the manufacturing process optimization. Optical measurement techniques provide noncontact, fast, sensitive, and accurate measurements assisting microfabrication.

Vibration modes of microdevices can be characterized using scanning laser Doppler vibrometry [1, 2], which involves moving either the device or the laser beam across its surface. Pointwise Doppler methods provide high accuracy at individual measurement points but necessitate extensive spatial scanning, significantly increasing measurement duration and complexity when assessing whole surfaces. Alternatively, full-field interferometric methods, operating in either stroboscopic [3-5] or time-averaging mode [5-8], are also commonly employed. Stroboscopic methods rely on synchronizing short illumination pulses with periodic motion, capturing distinct vibration states sequentially. Despite their high resolution, stroboscopic techniques typically require precise temporal synchronization and multiple measurements to reconstruct full-field data. Time-averaged interferometry (TAI) remains particularly advantageous due to its capability of visualizing micromechanical device vibration modes at video-rate speeds and with high spatial resolution, independent of the vibration frequency.

In general, in the case of interferometry, the measurand is not received directly from the measurement but appears spatially encoded in the form of a fringe pattern (e.g., interferogram, hologram, moiregram or besselogram). In the majority of cases, useful information is encoded in the phase distribution (optical path difference) of the recorded intensity image. For the TAI-based measurement the situation is different. Measurand (information about harmonic motion) is encoded in the amplitude modulation of the recorded interferogram. Considering the case of the vibrating object (harmonic sinusoidal vibration) and its TAI measurement, the interferogram intensity distribution can be described as:

$$I_{vibr}(x,y) = K(x,y)\left\{1 + C_{stat}(x,y)J_0\left(\frac{4\pi}{\lambda}a_0(x,y)\right)\cos(\varphi_{vibr}(x,y))\right\}, \quad (2)$$

where $K(x,y)$ describes background intensity, $C_{stat}(x,y)$ describes fringes contrast distribution, $\varphi_{vibr}(x,y)$ encodes the information about average position of the vibration object, $J_0$ denotes Bessel function of zero order (first kind) and its argument $a_0(x,y)$ is vibration amplitude distribution – the information we are interested in during the measurement.

For that reason, the full analysis of TAI-based interferogram, resulting in information retrieval (map of the vibration amplitude), can be divided into two steps, as shown in Fig.1. Firstly, we need to compute the fringe pattern amplitude modulation from the recorded interferogram generating the absolute value of the Bessel function - besselogram. Secondly, we need to estimate the argument of the Bessel function – besselogram phase distribution – to access the vibration amplitude distribution. One possible solution is the usage of the temporal phase-shifting algorithm (TPS) [9,10], which is a very accurate method but requires multiple frames recording. The issue is even more relevant in the case of time-averaging interferometry than in the case of classical interferometry, since it requires not only multiple time-shifted recordings but also precise modulation of the reference beam at the same frequency as the vibrating object, along with controlled phase shifts of this modulation. Those conditions demand the use of expensive modulators and result in a more complex, bulky, and costly experimental setup. The minimum number of frames needed (for 3-frames approach) considering 2-steps TAI analysis is 9 (3 interferograms piezo-phase-shifted to calculate single besselogram, and then 3 besselograms phase shifted by a reference arm modulator are needed to access the Bessel phase). Some attempts at single-frame algorithms for modulation determination have also been reported [11-14]. However, it is important to emphasize that all classical state-of-the-art algorithms (multi and single-frame) are designed to fit to cosine phase variations and generate fundamental errors due to mismatches with the Bessel function. Comparing Bessel and cosine functions, there is a discrepancy of the extrema positions of both functions, especially for the small argument values. Moreover, the zeros of the Bessel function are not equally spaced. In consequence, if in the argument estimation process the Bessel function is treated as cosinusoidal, the fringe like error can be observed in the resultant distribution of the vibration amplitude. It can be clearly noticed in Figs. 1(h) and 1(j).

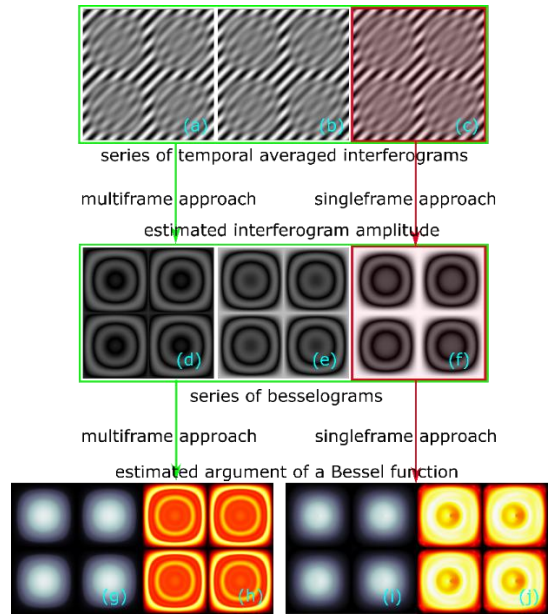

Fig. 1. The scheme of TAI interferogram analysis: (a)-(c) phase-shifted series of temporal averaged interferograms, (d)-(f) phase-shifted series of besselograms, absolute value of vibration amplitude and error maps for (g), (h) multi-frame approach (TPS) and (i), (j) single-frame approach (Hilbert spiral transform).

As an alternative to classical approaches, full-field optical measurement solutions based on convolutional neural networks (CNN) and deep learning are currently under rapid development. They have already been successfully applied to different tasks in optical metrology [15-36]. In this paper, we decided to use a deep learning approach called DeepBessel to leverage the capabilities of convolutional neural networks to aid the measurand estimation process in full-field vibration profilometry. To overcome the problem of incompatibility of the Bessel function with the classical fringe pattern analysis algorithm a CNN architecture was trained with the data input defined as simulated Bessel function and data output as absolute value of underlying argument encoded in a besselogram (see scheme presented in Fig. 2). The neural network architecture was inspired by the work [16]. The motivation for using deep learning for besselogram analysis is that even though the Bessel function cannot be simply described analytically, a convolutional neural network in the training process should be able to achieve an accurate estimation of the Bessel function argument. Using a deep learning approach there is no need for further correction procedure in estimated phase values.

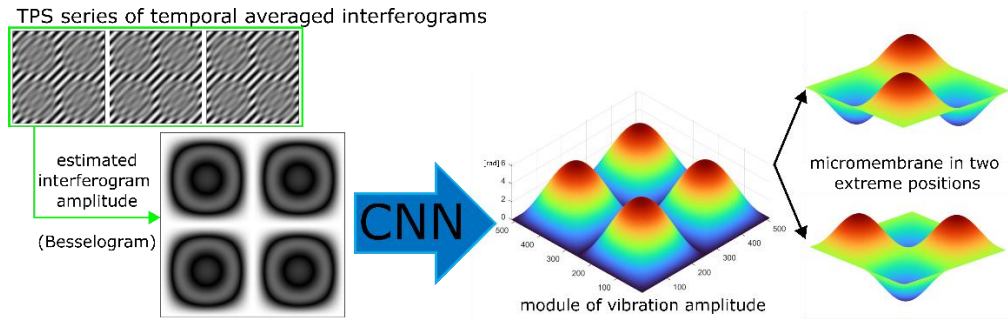

Fig. 2. Training scheme of deep learning solution proposed in this paper.

The paper is structured as follows: Section 2 introduces proposed DeepBessel neural network architecture and specificity of its training process; Section 3 compares proposed novel neural network technique with classical, cosine-based solutions using simulated and experimental data; Section 4 concludes the paper.

## 2. Estimation of the argument of Bessel function using neural network

To start the description of the training process we would like to introduce the generation process of training dataset. The input data of the neural network were synthetically generated besselograms. The main reasoning behind using deep learning for besselogram analysis was the fact that in the case of Bessel function there is a lack of clear analytical solution to the inverse problem (Bessel phase demodulation). In the case of classical vibration amplitude demodulation solutions in TAI [9-14] Bessel function is approximated by cosine. However, comparing Bessel function and cosinusoidal function, one may clearly see the discrepancy of the extreme positions of both functions, especially for the small argument values (see Fig.3). Moreover, the zeros of Bessel function are not equally spaced. In this work, because the training is based on simulations, we can train neural network to map the basic relationship between numerically generated Bessel function and its ground-truth argument (phase map which in experimental reality encodes the modulus of vibration amplitude distribution). In addition, the generation of synthetic images is much easier and less time-consuming than obtaining experimental data. The output data of the network was defined as the modulus of the vibration amplitude distribution (the argument of the first kind, zero-order Bessel function). In order to clearly illustrate the process of training set generation, a scheme was created and depicted in Fig. 4. One can notice the effects of the various operations described further in this section.

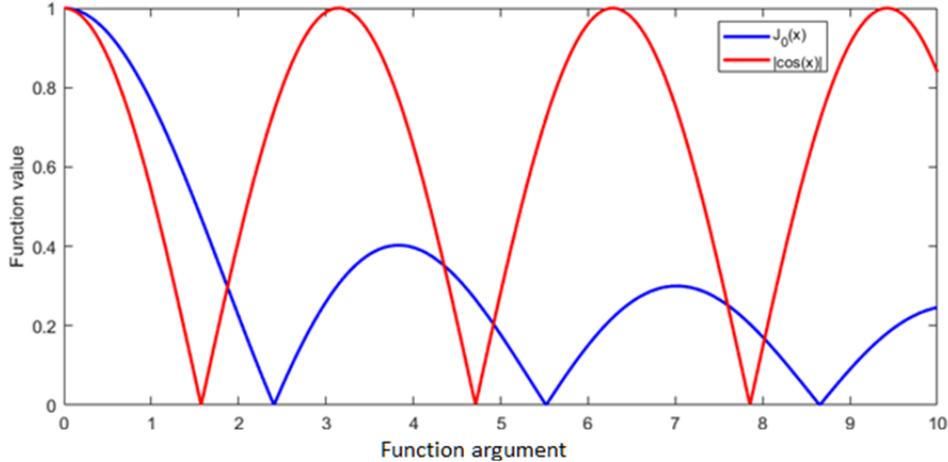

Fig. 3. Bessel and cosine functions

The training set generation process began by simulating the harmonic vibration of a square-shaped membrane. Its swing out of the equilibrium is represented by the expression:

$$u(x, y, t) = X(x)\, Y(y)\, T(t), \tag{3}$$

where the x, y are spatial coordinates in the plane of the membrane surface, the parameter t determines the time course. Functions X, Y and T are formed by applying the method of separation of variables from the wave equation. These are the eigenfunctions corresponding to the individual components of the variables (spatial and temporal). After superimposing the boundary conditions (determining the dimension of the membrane) and the initial conditions

(the course of the shape function and the velocity at the initial position), three component functions were obtained:

$$X_m = \sin\left(\frac{m\pi}{L_x}x\right), \quad (4)$$

$$Y_n = \sin\left(\frac{n\pi}{L_y}y\right), \quad (5)$$

$$T_{mn} = A\cos(\lambda_{mn}t) + A^*\sin(\lambda_{mn}t), \quad (6)$$

where the values m and n are the number of quantum vibration modes of the membrane and $L_x$, $L_y$ are the dimensions of the membrane. The value labelled $\lambda$ represents the frequency of the vibration. The variable A and its conjugate depend on the initial properties of the membrane. If we assume the time value at the initial moment, t = 0, an expression A* is zeroed out. Finally, the expression used to generate the membrane vibration amplitude distribution is as follows:

$$u(x, y, t) = \psi_{mn}(x, y) = A\sin\left(\frac{m\pi}{L_x}x\right)\sin\left(\frac{n\pi}{L_y}y\right), \quad (7)$$

where the parameter A defines the vibration amplitude maximum value. The parameters $Lx$ and $Ly$ correspond to the size of the membrane in pixels. As it has already been mentioned the output of proposed DeepBessel neural network is expressed as modulus of the vibration amplitude map (abs($\psi_{mn}(x, y)$)). The input of the neural network, in the form of the Bessel function, shows the same characteristics for both an argument represented by a concave (negative vibration amplitude) and a convex (positive vibration amplitude) function. In the neural network training process, connecting the vibration modes in different directions with same Bessel pattern would be misleading and would not allow the optimizing procedure to converge to the reasonable level of generalization.

In general, the frequency of vibration mode can be determined based on the following equation:

$$\lambda_{mn} = \frac{1}{2}\sqrt{\left(\frac{m}{L_x}\right)^2 + \left(\frac{n}{L_y}\right)^2}. \quad (8)$$

With the knowledge of the frequency values, it was possible to enrich the training set with superpositions of two different quantum modes. The necessary condition is the same value of the frequency $\lambda_{m_1n_1} = \lambda_{m_2n_2}$ of these modes. The superposition of two vibration modes was defined as follows:

$$\psi = a\psi_{m_1n_1} + b\psi_{m_2n_2}, \quad (9)$$

where in our simulation parameters α and β are random multipliers from an arbitrary range, in our case from 0 to 5.

Finally, based on the synthetically created vibration amplitude distributions, the Bessel function of the first kind of zero order $J_0(\psi)$ is calculated. As a result, a synthetic besselogram is created which, together with the absolute value of the corresponding vibration amplitude distribution, forms the pair (input and expected output) used for the DeepBessel network training process.

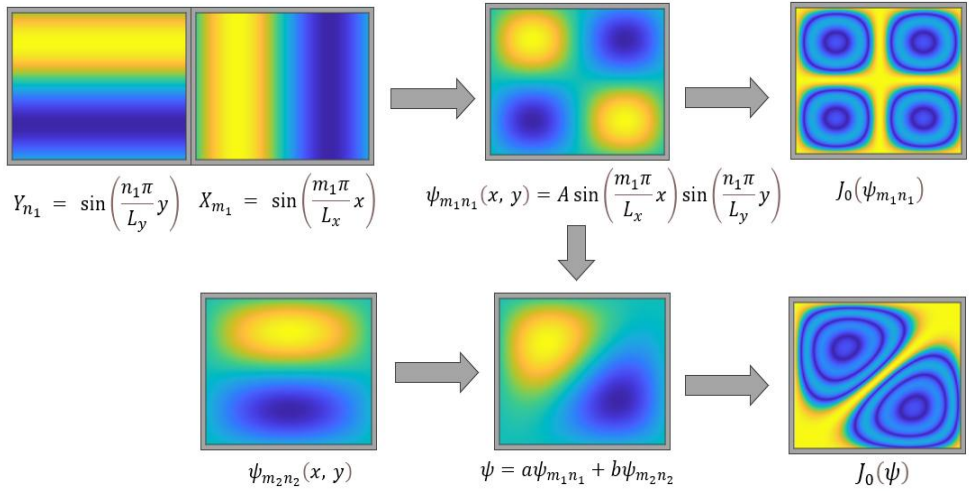

Fig. 4. Workflow of synthetic besselogram generation process, given by the example $m_1, n_1, n_2 = 2$ and $m_2 = 1$.

## *2.1. Simulated dataset for DeepBessel training process*

Preparation of a suitable training dataset is one of the key steps that needs to be performed in order to carry out a successful neural network training process. The training set should be characterized by a significant amount of unique data allowing to achieve the proper generalization of the trained neural network. It is often assumed that the minimum amount of training data must exceed several thousand data instances. In the case of training set used to train the DeepBessel neural network, the amount of data was decreased compared to the suggested standards, as it contains 1140 image pairs. The significant reduction in the amount of data for the training set was possible due to the local self-similarity property of fringe patterns under study - besselograms. Even if the underlying besselogram argument varies the fringes themselves have a very similar structure making the optimization process easier. Additional advantage is that a smaller dataset contributes to a shorter training process.

The training set was defined with the use of the previously discussed method of obtaining a synthetic vibration amplitude distribution (Eq. (3-7)) assuming the size of the membrane as $L_x = L_y = 512\ px$. The variable simulation parameters were vibration amplitude value (A) and number of quantum modes (m, n). The range of the first parameter was changed from 1 to 15 radians with a step value of one radian. The second parameter, the number of quantum vibration modes in both x and y planes (value m and n), was changed from 1 to 4 with a step value of one. The above values correspond to the ranges of vibrations often achieved in experimental studies. By imposing different parameter values, it was possible to generate an extended training set. In Fig. 5 examples of synthetically generated besselograms are presented. They are divided in terms of the number of vibration modes: m=1 n=1, m=2 n=2, m=4 n=4 (11, 22, 44) and amplitude variations: 1, 5, 10, 15 radians.

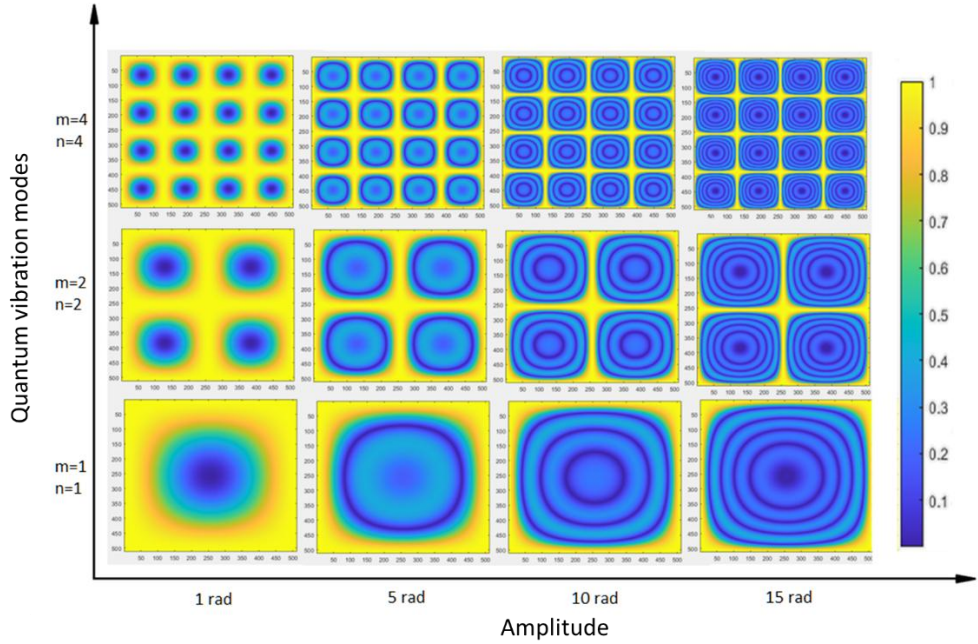

Fig. 5. Exemplary synthetically generated besselograms from training dataset with different vibration amplitude values (in radians), and the number of quantum vibration modes (m, n).

For the training purpose, the dataset was randomly divided into two subgroups: training and validation datasets. The first counted 1000 images, which were used to train the neural network. The remaining images were used to verify the effectiveness of the neural network's prediction accuracy at different stages of the training process. The minibatch size was set as 1 and whole training lasted 40 epochs with the use of shuffle after each epoch. Adam optimization algorithm was used, and the initial learning rate parameter has been set to $10^{-4}$. The training process was performed with single GPU on a NVIDIA GeForce GTX 1080 8GB. The detailed information about the training time and single data execution time for each of analyzed networks within this work is presented in Table 1, where DeepBessel architecture parameters are bolded.

Table. 1. Summary of learning times for a neural network with different parameters and their prediction times for a single data frame.

| Network parameters: | training time [min] | single data execute time [s] |
|---|---|---|
| 50 filters, 3 paths | 517 | 0.166917 |
| 50 filters, 4 paths | 703 | 0.180583 |
| **50 filters, 5 paths** | **848** | **0.229406** |
| 50 filters, 6 paths | 941 | 0.233467 |
| 50 filters, 7 paths | 1071 | 0.257612 |
| 20 filters, 5 paths | 336 | 0.109153 |
| 30 filters, 5 paths | 523 | 0.145311 |
| 40 filters, 5 paths | 682 | 0.180511 |
| 60 filters, 5 paths | 1024 | 0.278074 |

## 2.2. DeepBessel network architecture

The neural network architecture developed for besselogram analysis focuses on the concept of using a convolutional layers structure (CNN) in combination with residual layers and implementation of several separate training paths in parallel. Mentioned network architecture has already been successfully applied to a few different optical metrology tasks and we adapted it here to besselogram analysis [16, 34-36]. Each path differs in the size of the filter mask used in the downsampling operation. For clarity the proposed neural network architecture has been divided into three steps and presented in the form of a simplified block diagram in Fig. 6.

The size of the images in the training dataset was set to be 512 x 512 pixels (size of a simulated membrane). Nevertheless, it should be noted that the proposed network (after the training process is finished) can process any image sizes, which is a significant advantage over other popular encoder-decoder-based network architectures, e.g., UNet [37]. This flexibility is achieved due to the consistence of the matrix size in each of the network paths.

The initial part of the network architecture relates to the separation of the input image into various paths, Fig. 6 (Initialization). Each successive path is characterized by a larger filter mask for the downsampling operation. This approach translates into a reduction of the image dimension relative to each successive path, allowing the training to focus on different image details. The number of paths is one of the parameters to tune while searching for proper network architecture. Each of the paths consist of convolutional layers with a filter size of 3x3 and padding set to prevent changes in the dimensions of the output images. The number of convolution filters in each layer is a second parameter to tune for defining the final network architecture. Activation function in each of convolutional layers was set as ReLU function. Such pair is a standard connection in contemporary deep learning models.

The next step focuses on four repeating residual blocks, the same number in each path, Fig. 6 (Processing). At the beginning of each path (excluding the first one) the downsampling layer is introduced, where the input image dimensions are reduced by applying a scale factor. The downsampling procedure is determined using bilinear interpolation. The residual blocks were used to prevent the vanishing gradient during training, which may occur in deep neural network architectures.

The last step begins with the upsampling procedure to increase the dimensionality of the processed feature maps (outputs from the last residual block in each path) to the same size as the input image, Fig. 6 (Finalization). The upsampling is followed by the three convolution layers to average and smooth out estimated maps. The final step is the concatenation of all the individual channels of the network, that is, the feature maps obtained from each path. This results in the fusion of various relevant image features. After this operation, we get a single feature map, which is passed to the last convolutional layer defining the final output of the trained network.

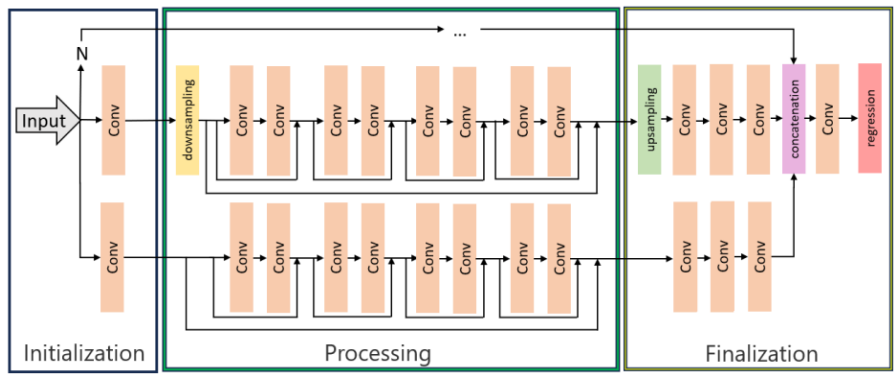

Fig. 6. Simplified block diagram of the proposed neural network architecture, where the value N corresponds to the number of training paths.

## 3. Experiments and results

In this section the numerical evaluation of the developed DeepBessel network and fine-tuning the architecture parameters (number of paths and convolutional filters) is presented. The main goal was maximizing the generalization of estimated solutions. Final network architecture is tested with the use of experimentally recorded besselograms and compared to state-of-the-art single-frame besselogram phase demodulation solution based on Hilbert spiral transform [14].

### 3.1. Numerical evaluation and synthetic data processing

Based on the analysis attached in the appendix A it can be concluded that the right choice for DeepBessel is to use a neural network architecture with 50 filters in each convolutional layer and 5 parallel training paths. The choice was made based on achieved low prediction errors in comparison to other analyzed combinations and the highest stability, with no noticeable spikes in prediction errors.

To test the predictive capabilities of the proposed neural network architecture the mean square error (MSE) was calculated. In this section we explore DeepBessel accuracy using synthetic data not included in the training dataset. The dataset used in this study was simulated as it is described in Section 2. By applying equation (9), the superposition of two different vibration modes with different amplitude values was calculated. In addition, for each superposition vibration modes were multiplied by a random constant in the range from 0 to 5. A study was carried out on the resulting dataset, examples of which, along with their generation parameters, are presented below in Fig. 7 and 8. Additionally, obtained results are compared with the single-frame state-of-the-art solution called Hilbert spiral transform (HST) [14].

In the case of first exemplifying besselogram (Fig. 7), the vibration amplitude distribution prediction estimated by DeepBessel was closely related to the expected results (mean square error was 0.0206), as shown by the minor global differences in the associated error map (Fig. 7(c)). The largest error values can be seen around the maximum amplitude, but still they are preserved on an acceptable level. Additionally, no systematic, fringe-like error is visible. In the case of the reference HST method, the presence of fringe-like error can be seen clearly on the error map (Fig. 7(f)), which indicates a non-matching of the HST approach for the Bessel function. Mean square error was 0.8097 which is much higher value than the neural network results. In addition, globally the amplitude differences are greater than for the DeepBessel with the significantly larger local peaks of error values.

Similar conclusions can be drawn for the second presented case (Fig. 8). The prediction estimated by the DeepBessel is far more accurate than the one obtained by the classical method. MSE for the network prediction was 0.0120 while for the classical method it was 0.8916. Even though locally occurring peaks in the DeepBessel estimated amplitude error map (Fig. 8(c)) are slightly higher than in the previous case (Fig. 8(c)) they are still significantly lower than for a classical solution. Once again, for the classical method, significant errors connected with the mismatch of the cosine function and the Bessel function can be seen.

Additional analysis based on simulated data is presented in appendix B, where we have provided the comparison between our DeepBessel method and classical single- and multi-frame approaches.

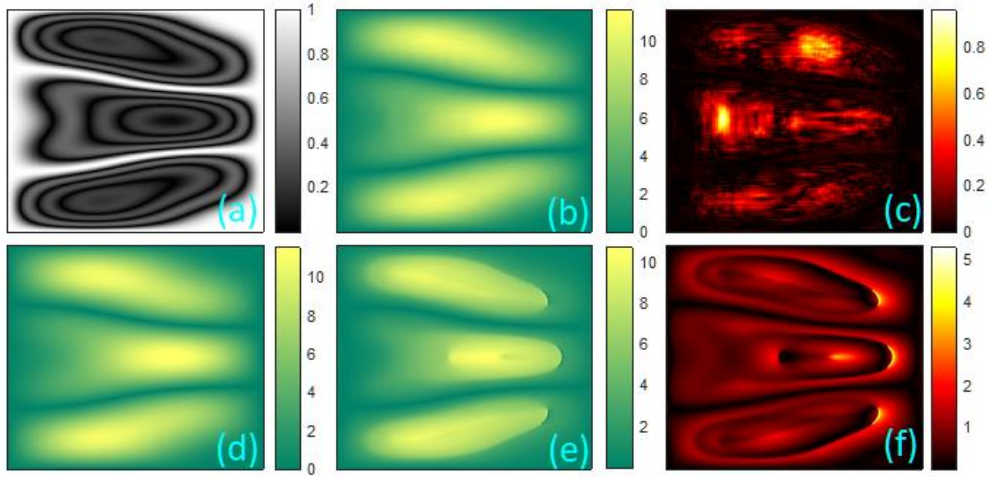

Fig. 7. Analysis of predictive accuracy of DeepBessel network compared to the classical method: (a) synthetically generated besselogram as superposition of two vibration modes with parameters: 1,3 and 2,5 and amplitudes of 2.7 and 1.9 radians with random multipliers: 3.72 and 1.73, (b) DeepBessel prediction, (c) difference of network prediction from expected result, (d) expected vibration amplitude distribution, (e) HST method prediction, (f) difference of HST method prediction from expected result.

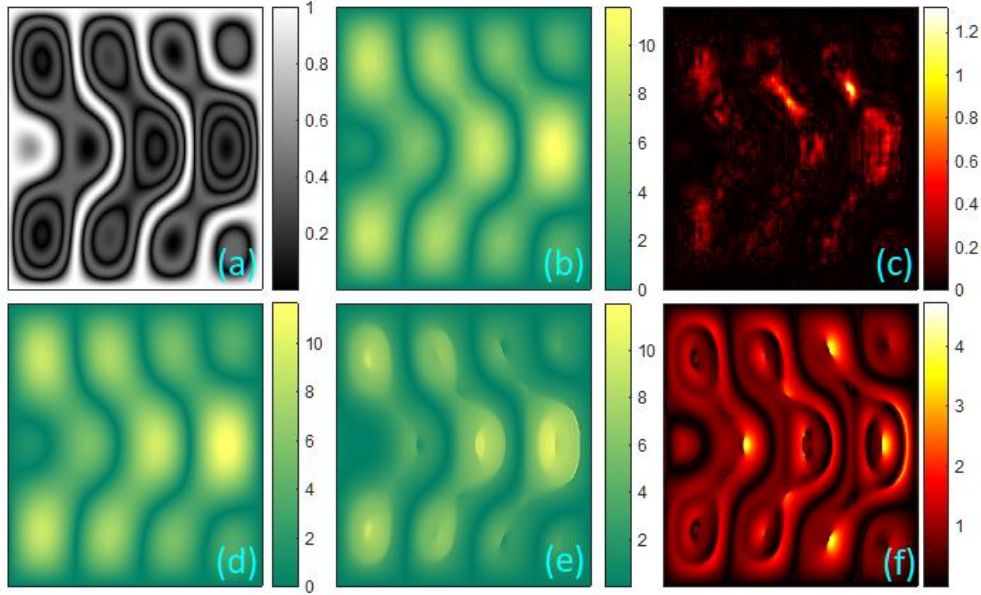

Fig. 8. Analysis of predictive accuracy of DeepBessel network compared to the classical method: (a) synthetically generated besselogram as superposition of two vibration modes with parameters: 4,3 and 3,1 and amplitudes of 4.1 and 3.2 radians with random multipliers: 1.53 and 1.78, (b) DeepBessel prediction, (c) difference of network prediction from expected result, (d) expected vibration amplitude distribution, (e) HST method prediction, (f) difference of HST method prediction from expected result.

### 3.3. Experimental data processing

This section discusses the predictive capabilities of the proposed DeepBessel neural network using experimentally recorded data [7]. In time-averaged interferometry, obtaining experimental ground truth for vibration amplitude typically requires multi-frame phase-shifting or stroboscopic measurements synchronized with the vibration cycle. Such approaches significantly increase experimental complexity and measurement time and contradict the single-shot nature and motivation of the proposed method. For that reason, as the reference estimated results are compared to the state-of-the-art single-shot method - Hilbert spiral transform [14]. Since DeepBessel was trained using synthetically generated besselograms it is important to preprocess the experimentally recorded data before feeding it to the neural network. To preserve a fair comparison between two methods, it was decided to filter the input images to minimize noise using block-matching 3D denoising (BM3D) algorithm [38]. Noisy data was not included in the training dataset to not bias our solution towards any specific noise type. Additionally, the background in input images for the HST method was filtered using unsupervised variational image decomposition (uVID) algorithm [39]. This additional filtration step for classical approach was done to slightly change the structure of the besselogram oscillations and match the cosine function more closely (bringing mean value around zero and even out contrast). In the case of a neural network, this operation was not needed, moreover it could have hindered its prediction, for which reason this step was omitted. After the prefiltration of the experimental images, a comparative study of the two methods was carried out, the result of which is presented in Figs. 9 and 10.

In the first example (Fig. 9), it can be seen that both methods managed to determine the distribution of the vibration amplitude. However, the prediction result in the case of classical approach is corrupted by previously described, characteristic fringe-like error. This is particularly evident observing the additionally included difference map.

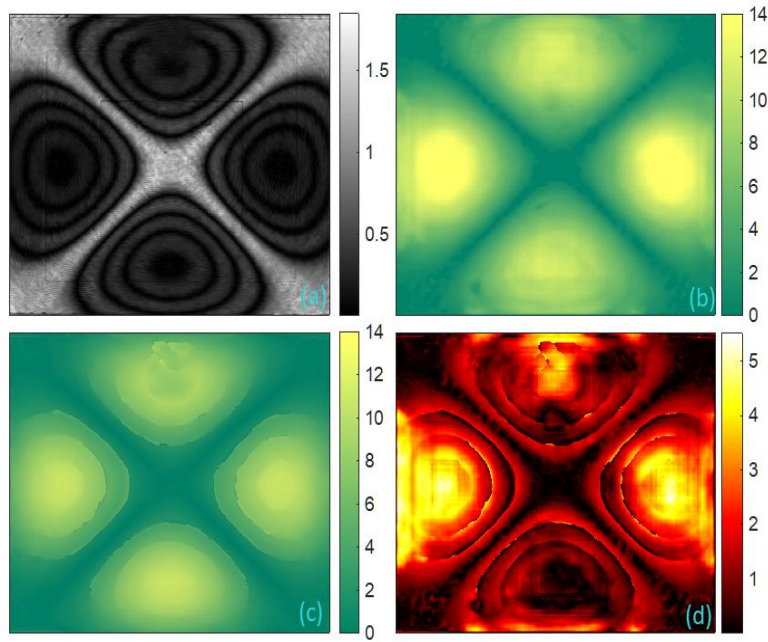

Fig. 9. Experimentally recorded data: (a) besselogram obtained from experimental data for a 172 kHz oscillating membrane, (b) prediction of the DeepBessel neural network, (c) prediction of HST method, (d) difference between the results of the analyzed methods.

In Fig. 10 besselogram is described by only one fringe with an increase in value at the characteristic point of their connection. In this case, the poor result obtained by the HST method is subject to significant errors connected also with a phase unwrapping procedure. Additionally, in this case, the neural network managed to determine the amplitude of the vibration distribution with clearly distinguishable maximum where the classical method was not able to generate any meaningful result.

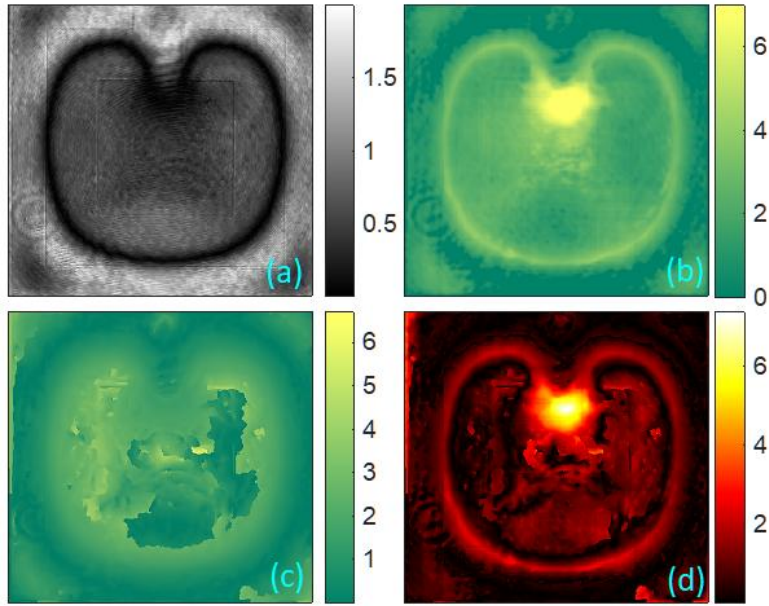

Fig. 10. Experimentally recorded data: (a) besselogram obtained from experimental data for a 100 kHz oscillating membrane, (b) prediction analysis of the neural network DeepBessel, (c) prediction analysis of HST method, (d) differences between the results of the methods used.

Stroboscopic Interferometry (SI) [40] was additionally considered as an experimental reference technique, since it enables direct reconstruction of the instantaneous out-of-plane displacement of a vibrating object. In SI, the object is illuminated with short optical pulses that are synchronized with the excitation signal, typically with pulse durations below 2% of the vibration period. This pulsed illumination "freezes" the mechanical motion, allowing the interferometer to capture a sequence of quasi-static fringe patterns corresponding to well-defined time samples of the vibration cycle. By acquiring interferograms at multiple temporal delays and subtracting them from the static shape, the transient displacement and maximum vibration amplitude can be retrieved without relying on Bessel-function inversion. As demonstrated experimentally in [41], SI provides highly accurate amplitude maps. For this reason, in our comparative study SI serves as a high-quality experimental benchmark for vibration amplitude map reconstruction, complementing the synthetic ground truth used within this paper for quantitative evaluation. The estimated results for SI, DeepBessel and HST are presented in Fig. 11. DeepBessel (Fig. 11(b)) yields a reconstruction closely matching the stroboscopic reference (RMSE=0.7497 rad). In contrast, the HST method (Fig. 11(c)) exhibits significantly higher errors (RMSE=2.1417 rad).

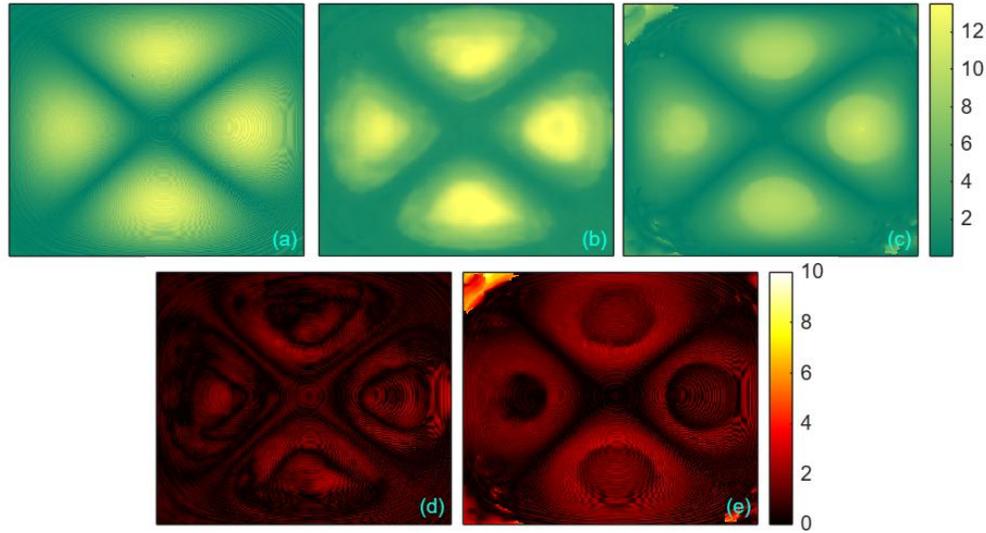

Fig. 11. Comparison of vibration amplitude reconstruction methods using experimental data: (a) stroboscopic interferometry result, (b) DeepBessel prediction, (c) HST prediction, (d) map difference between reference (a) and (b), (e) map difference between reference (a) and (c).

## 4. Conclusions

In this paper, we introduced DeepBessel, a deep learning-based approach tailored for besselogram analysis in full-field vibration profilometry. By leveraging convolutional neural networks (CNNs), DeepBessel directly estimates the vibration amplitude distribution from a single besselogram (calculated accessing the amplitude term of phase-shifted time-averaged interferograms sequence).

Compared to classical single-frame approaches, such as the Hilbert Spiral Transform (HST) method [14], DeepBessel offers several significant advantages. Firstly, it effectively mitigates errors caused by the mismatch between the Bessel function (specific for temporal averaging interferometry analysis) and cosine-based phase demodulation techniques, which are commonly used in interference-based optical metrology. Secondly, DeepBessel operates with high precision on both synthetic and experimental data, significantly reducing reconstruction errors while preserving the fine details of the vibration amplitude distribution. The experimental validation confirmed that our approach outperforms traditional methods.

Additionally, DeepBessel benefits from its ability to process single-frame besselograms without requiring multiple phase-shifted recordings, significantly reducing the measurement cost. The proposed network architecture efficiently generalizes across a wide range of vibration patterns, demonstrating robustness against variations in vibration amplitude values and modes frequency. Furthermore, the deep learning model does not require additional postprocessing correction steps, simplifying the measurement pipeline and enhancing the reliability of vibration profilometry.

The results of this study highlight the strong potential of deep learning in optical metrology, specifically by addressing challenges associated with full-field vibration measurements. The successful implementation of DeepBessel enables more accurate and automated analysis of microelectromechanical systems (MEMS/MOEMS) vibrations.

**Funding**

This work has been funded by the National Science Center Poland (PRELUDIUM 2021/41/N/ST7/04057, OPUS 2024/55/B/ST7/02085) and the Warsaw University of Technology within the Excellence Initiative: Research University programme (YOUNG PW 504/04496/1143/45.010012).

**Disclosures**

The author declares no conflicts of interest.

**Data Availability.**

Data may be obtained from the authors upon reasonable request. The trained DeepBessel model is made freely available on GitHub https://github.com/MariaSi1/DeepBessel.

**Appendix A.**

For the evaluation purpose, synthetical besselograms of different amplitude values and quantum vibration modes were generated. The test of the DeepBessel network was performed on two separate subsets of training datasets. The first dataset was built up using besselograms generated with fixed vibration mode (m=2 and n=2) and amplitude values changed in the range from 1 to 15 radians with 1 radian increments. The second dataset consisted of besselograms generated with a fixed amplitude of 10 radians and the number of quantum oscillation modes as the variable values. The parameters m and n were selected from 1 to 4 for all possible combinations. Both datasets were generated in order to distinguish the influence of changing the amplitude value and quantum vibration mode number onto the estimated model accuracy.

The first tuned neural network architecture parameter was the number of filters in each of the convolutional layers. In Fig. A1 the MSE values were calculated for the dataset with changing vibration amplitude values. Five variants of the tuned neural network architecture were tested, each with a different number of filters in convolutional layers. For all estimated curves a strong trend of an increase in the error value for higher amplitude values can be noticed, especially evident for the network architectures with lower number of filters. With the increase of amplitude value, the complexity level of analyzed besselograms also increases in the meaning of visible fringes number. The mentioned variation in error plots estimated for 20, 30 and 40 filters may indicate that the complexity of network architecture was too low for proper fitting to the dataset (underfitting). By far the smallest prediction errors belonged to networks containing 50 or 60 filters. For smaller amplitude values (1-6 rad), the network marked in green (60 filters) had a smaller error. For larger amplitude values (10-13), the network marked in purple (50 filters) was more effective. It is worth noting, however, that the observed differences in the errors of networks characterized by 50 or 60 filtering operations in convolution layers are negligible. It means that further increasing of DeepBessel complexity level does not significantly improve the predictive capabilities. Further fine tuning to achieve optimal network generalization is going to be achieved by adjustment of number of paths in network architecture.

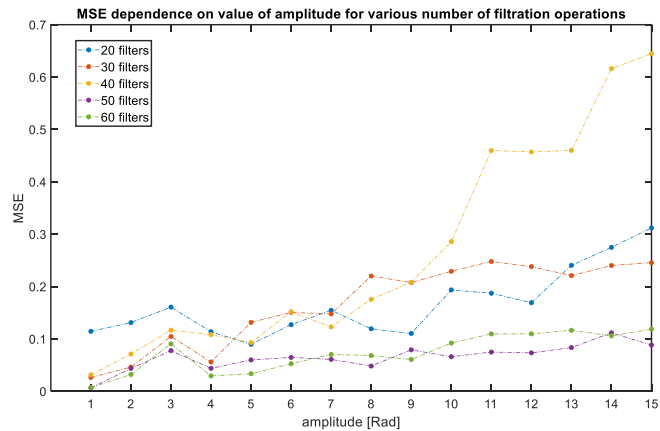

Fig. A1. The influence of vibration amplitude value on the obtained prediction error (RMS) for different number of filters in the designed neural network convolutional layers.

The next step was to analyze the predictive capabilities of the proposed neural network using dataset with variable number of quantum vibration modes. The graph presented in Fig. A2 illustrates the effect of changing the filtering operations number in convolutional layers on the prediction error. A clear separation can be observed in prediction efficiency. The first three networks (with filter numbers of 20, 30 and 40) have significantly higher errors than those with 50 and 60 filters. In this case, the neural network trained with 50 filters in convolutional layers has higher prediction efficiency. Connecting the conclusions drawn from the both studies (Fig. A1 and Fig. A2), the neural network with 50 filters in the convolutional layers was chosen as an optimal one for assumed besselogram analysis task.

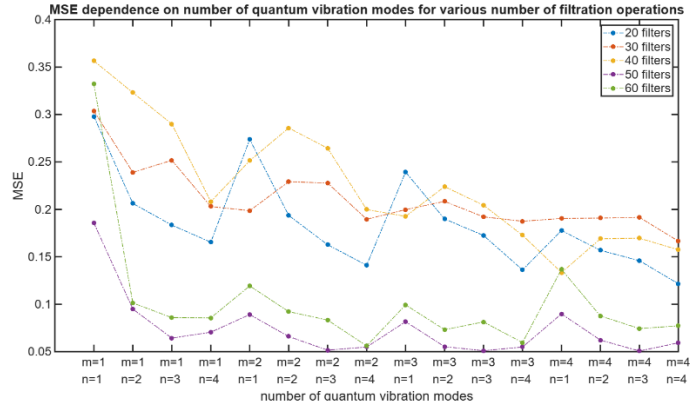

Fig. A2. The influence of the number of quantum vibration modes on the obtained prediction error (RMS) for different number of filters in the designed neural network convolutional layers.

Subsequently, in the same manner as described above, the impact of the number of parallel training paths on the neural network prediction accuracy was analysed (Fig. A3 and Fig. A4). The number of training paths varied in the range from 3 to 7. The filter number in the convolutional layers was fixed as previously estimated value of 50. Both in Fig. A3 and Fig. A4, it can be seen that the networks with low number of training paths (1 and 2) are associated with noticeable spikes in prediction error. As mentioned previously, the network inability of proper generalization and underfitting is clearly visible. In the case of 6 and 7 paths, a significant increase in prediction error can be observed for an amplitude value of three radians (Fig. A3). The most stable architecture, with the lowest prediction error values for all analyzed cases, turned out to be the neural network trained with 5 parallel training paths.

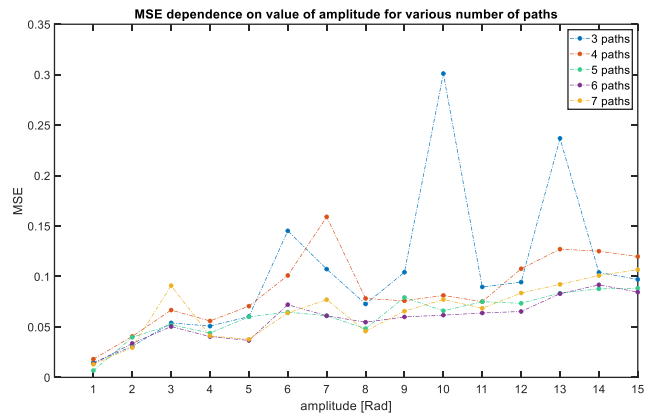

Fig. A3. The influence of vibration amplitude value on the obtained prediction error (RMS) for different numbers of parallel training paths in the designed neural network.

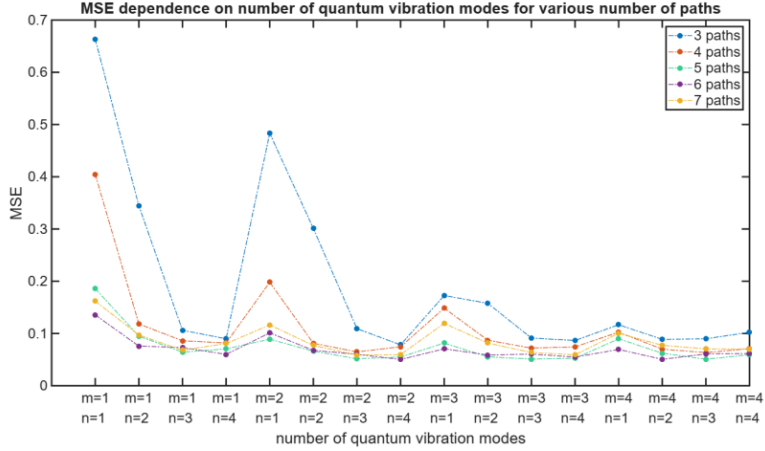

Fig. A4. The influence of the number of quantum vibration modes on the obtained prediction error (RMS) for different numbers of parallel training paths in the designed neural network.

**Appendix B.**

In order to provide a quantitative reference for evaluating the reconstruction accuracy, a comparative study was carried out using synthetically generated besselograms, for which the vibration amplitude distribution is exactly known. This allows objective error analysis with respect to ground truth, which is generally unavailable in single-shot experimental time-averaged interferometry.

The evaluated methods include the classical multi-frame Temporal Phase Shifting (TPS) approach, a single-frame Hilbert Spiral Transform (HST)-based method, and the proposed DeepBessel neural network (Figs. B1 and B2). TPS reconstruction was performed using three and five phase-shifted besselogram frames. Although TPS is commonly regarded as a reference technique in interferometric measurements, it is inherently optimized for cosine-modulated signals. As a consequence, when applied to besselograms it introduces systematic reconstruction errors, which remain visible even when the number of phase-shifted frames is increased. Moreover, the requirement for multiple frames significantly increases acquisition time.

The single-frame Hilbert spiral method alleviates the need for multiple acquisitions and shows improved performance compared to TPS, however, its accuracy is still limited by the analytical mismatch between the assumed cosine model and the actual Bessel-function modulation.

In contrast, the DeepBessel neural network achieves the highest reconstruction accuracy using only a single besselogram frame (see Table B1). By learning the direct mapping between the besselogram and its underlying argument, the proposed approach effectively mitigates errors related to model mismatch, providing improved robustness and precision compared to both classical multi-frame and single-frame techniques.

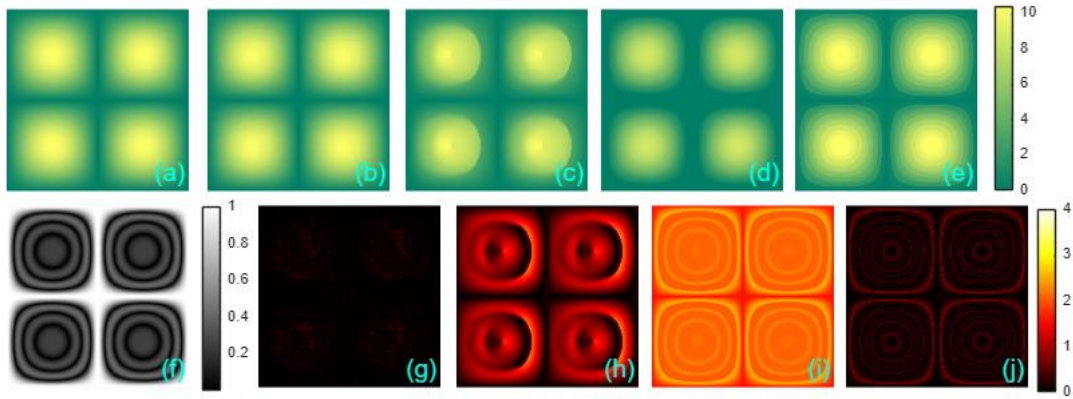

Fig. B1. Comparison of vibration amplitude reconstruction methods using synthetic data: (a) ground truth vibration amplitude distribution (m = 2, n = 2, amplitude 6 rad), (b) DeepBessel prediction, (c) HST prediction, (d) TPS prediction using three phase-shifted frames, (e) TPS prediction using five phase-shifted frames, (f) corresponding besselogram, difference maps between the ground truth (a) and results obtained with: (g) DeepBessel, (h) HST, (i) TPS (3 frames), and (j) TPS (5 frames).

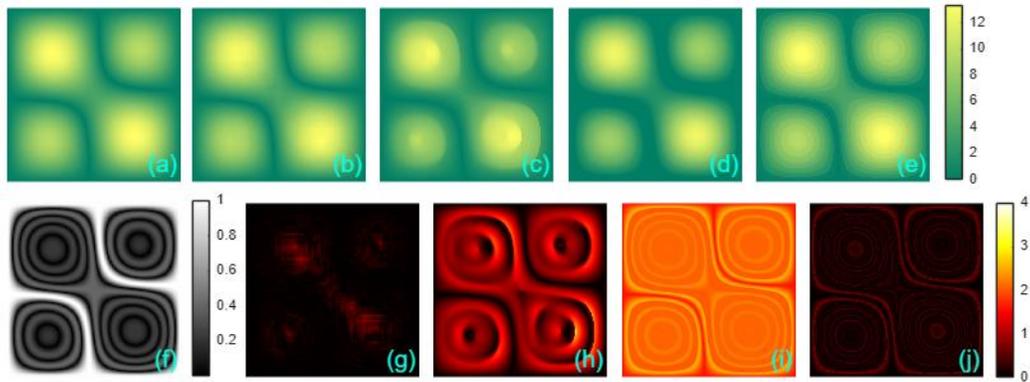

Fig. B2. Comparison of vibration amplitude reconstruction methods using synthetic data: (a) ground truth vibration amplitude distribution as superposition of two vibration modes (m1=1, n1=1 and m2=2, n2=2 and amplitudes of 1 and 7 radians with random multipliers: 3.65 and 1.81), (b) DeepBessel prediction, (c) HST prediction, (d) TPS prediction using three phase-shifted frames, (e) TPS prediction using five phase-shifted frames, (f) corresponding besselogram, difference maps between the ground truth (a) and results obtained with: (g) DeepBessel, (h) HST, (i) TPS (3 frames), and (j) TPS (5 frames).

Table. B1. Summary of Mean Squared Error (MSE) for different methods

| Method | MSE fig. B1 | MSE fig. B2 |
| --- | --- | --- |
| DeepBessel | 0.0038 | 0.0166 |
| HST | 0.7268 | 0.7413 |
| 3 frames TPS | 3.2590 | 3.4785 |
| 5 frames TPS | 0.0556 | 0.0530 |